\documentclass[12pt, draftcls, onecolumn, twoside]{IEEEtran}
\usepackage[utf8]{inputenc} 
\usepackage[T1]{fontenc}
\usepackage{url}
\usepackage{ifthen}
\usepackage{cite}
\usepackage[cmex10]{amsmath} 
\usepackage{amssymb}
\usepackage{graphicx}
\usepackage{multirow}
\usepackage{array}
\usepackage{makecell}
\usepackage{algorithm}
\usepackage{algpseudocode}
\usepackage{comment}
\usepackage[english]{babel}
\usepackage{romannum}
\usepackage{float}
\usepackage{tikz}
\usepackage{mathtools}

\usepackage{etoolbox}
\makeatletter
\patchcmd{\@makecaption}
{\scshape}
{}
{}
{}
\makeatother

\newtheorem{thm}{Theorem}
\newtheorem{lem}{Lemma}

\newtheorem{defn}{Definition}

\newtheorem{exmp}{Example}

\newtheorem{rem}{Remark}

\begin{document}
	\title{Improved Multi-access Coded Caching Schemes from Cross Resolvable Designs} 
	
	\author{%
		\IEEEauthorblockN{Pooja Nayak Muralidhar, Digvijay Katyal and B. Sundar Rajan}\\
		\IEEEauthorblockA{Department of Electrical Communication Engineering, Indian Institute of Science, Bengaluru 560012, KA, India \\
			E-mail: \{poojam,digvijayk,bsrajan\}@iisc.ac.in}		
	}
	
	\maketitle
	
\begin{abstract}
Recently multi-access coded caching schemes with number of users different from the number of caches  obtained from a special case of resolvable designs called Cross Resolvable Designs (CRDs) have been reported and a new performance metric called rate-per-user has been introduced \cite{KNRarXiv}. In this paper we present a generalization of this work resulting in multi-access coded caching schemes with improved rate-per-user.
\end{abstract}
\section{INTRODUCTION}	
Coded Caching has drawn considerable attention after the work of \cite{MaN}, which proposed a cache placement scheme
for uncoded content storage and a coded multicasting delivery
strategy that helped to provide both global and local caching gain for a significant delivery rate reduction. The main focus of coded caching is in designing schemes with reduced rate and practical subpacketization levels. Over the years there have been many approaches aimed at reducing subpacketization levels that involved developing coded caching schemes from resolvable designs from linear block codes \cite{TaR}, block designs\cite{KrP1}, placement delivery arrays \cite{YCTC}. While the focus has been mostly on setups with users equipped with dedicated caches, the setups where users have to share caches is also of utmost interest \cite{SBP2} \cite{AAA} \cite{BL}. Another interesting setup is where a user has access to multiple caches and vice versa. This setup is motivated from the fact that placing cache at local access points could significantly reduce the base station transmission rate, with each user being able to access content at multiple access points along with the base station broadcast.\\
The setting of multi-access setup was first considered in the work of \cite{HKD}, in which the setting where $K$ users and $K$ caches, where each user is associated to $z$ nearby caches in a cyclic way was dealt with. Later on, many schemes have been proposed for this multi-access setup\cite{SPE}\cite{RaK3}. 
The work of \cite{KNRarXiv} gives a multi-access setup through a combinatorial design called Cross Resolvable Designs (CRD), which was found to support a large number of users at low subpacketization levels. A metric called per-user-rate or rate per user was also introduced in \cite{KNRarXiv}, which allowed to compare different coded caching setups. The scheme proposed in \cite{KNRarXiv} was found to provide lower per-user-rates than Maddah Ali Niesen (MaN) scheme for the high memory regime.

\begin{figure}[H]
	\begin{center}
		\includegraphics[width=7cm,height=6cm]{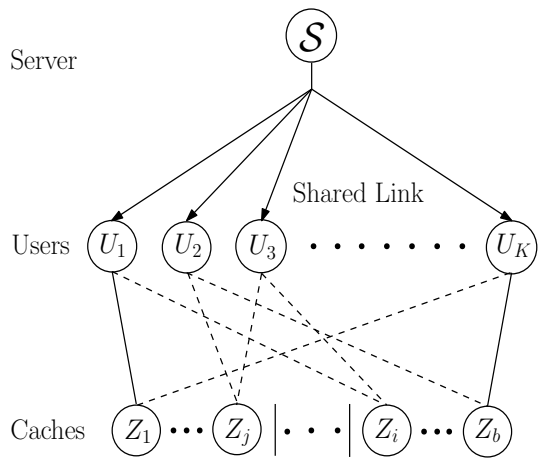}
		\caption {Problem setup for multi-access coded caching with $K$ users, $b$ caches and each user, connected to $z$ caches.}
		\label{fig1}
	\end{center}
\end{figure}

\subsection{Multi-access Coded Caching - System Model}
\label{sec1A}
Fig. \ref{fig1} shows a multi-access coded caching system with a unique server $\mathcal{S}$ storing $N$ files $W_{1}$,$W_{2}$,$W_{3}$,\dots,$W_{N}$ each of unit size. There are $K$ users in the network  connected via an error free shared link to the server $\mathcal{S}.$ The set of users is denoted by $\mathcal{K}.$  There are $b$ number of helper caches each of size $M$ files. Each user has access to $z$ out of the $b$ helper caches. Let $\mathcal{Z}_k$ denote the content in the $k$-th cache. It is  assumed that each user has an unlimited capacity link to the caches it is connected to. 

The setup where users are equipped with dedicated caches can be viewed as a special case of the scheme corresponding to Fig. \ref{fig1} with $b=K$ and $z=1$.

In any coded caching scheme there are two phases: the placement phase and the delivery phase. During the placement phase certain parts of each file are stored in each cache which is carried out during the off-peak hours. During the peak hours each user demands a file and the server broadcasts  coded transmissions such that each user can recover its demand by combining the received transmissions with what has been stored in the caches it has access to. This is called delivery phase. The coded caching problem is to jointly design the placement and the delivery with minimal number of transmissions to satisfy the demands of all the users. The amount of transmissions used in the unit of files is called the {\it rate} or the {\it delivery time}. Subpacketization level is the number of packets that a file is divided into. Coding gain is defined as the number of users benefited in a single transmission. 

\subsection{Contributions}
\label{contrib}
This work is a generalization of the multi-access scheme in \cite{KNRarXiv}. A modification in the set-up of \cite{KNRarXiv} is introduced  which enables having a large number of users than in \cite{KNRarXiv}. This allows to achieve lower per-user-rates than \cite{KNRarXiv} at the same subpacketization levels.
\subsection{Preliminaries}
\label{Prelim} 
In this section, we review some of the definitions in \cite{Stinson} and \cite{KNRarXiv}. We use a class of combinatorial designs called resolvable designs\cite{Stinson} to specify placement in the caches.

\begin{defn}\cite{TaR}
	A design is a pair  $(X, \mathcal{A})$ such that
	\begin{itemize}
		\item $X$ is a finite set of elements called points, and
		\item $\mathcal{A}$ is a collection of nonempty subsets of $X$ called blocks, where each block contains the same number of points.
	\end{itemize}
\end{defn}

\begin{defn}\cite{TaR}
	A parallel class $\mathcal{P}$ in a design $(X, \mathcal{A})$ is a subset of disjoint blocks from $\mathcal{A}$ whose union is $X$. A partition of $\mathcal{A}$ into several parallel classes is called a resolution, and $(X, \mathcal{A})$ is said to be a resolvable design if A has at least one resolution.
\end{defn}

\begin{exmp}
\label{exmp1}
Consider a design specified as follows.
\begin{align*}
X =& \{1, 2, 3, 4\}, \text{    and   } \mathcal{A} = \{\{1, 2\},\{1, 3\},\{1, 4\},\{2, 3\},\{2, 4\},\{3, 4\}\}.
\end{align*}
It can be observed that this design is resolvable with the following parallel classes.
\begin{align*}
\mathcal{P}_1 =& \{\{1, 2\},\{3, 4\}\}, ~~ 
\mathcal{P}_2 = \{\{1, 3\},\{2, 4\}\}, \text{   and   } 
\mathcal{P}_3 = \{\{1, 4\},\{2, 3\}\}.
\end{align*}
	
Note that in above example, $\mathcal{P}_1$, $\mathcal{P}_2$, $\mathcal{P}_3$ forms a partition of $\mathcal{A}$. If $\mathcal{A}$ = \{\{1, 2\},\{1, 3\},\{3, 4\},\{2, 4\}\}, we get another resolvable design with two parallel classes $\mathcal{P}_1$ and $\mathcal{P}_2$.
\end{exmp}

\begin{exmp}
\label{exmp2}
Consider a design specified as follows.
\begin{align*}
X =& \{1, 2, 3, 4, 5, 6\}, \text{ and  } \mathcal{A} = \{\{1, 2 ,3\},\{4, 5, 6\},\{1, 4 ,5\},\{2, 3, 6\}\}.
\end{align*}
It can be observed that this design is resolvable with the following parallel classes.
\begin{align*}
\mathcal{P}_1 =&\{\{1, 2 ,3\},\{4, 5, 6\}\} \text{  and } 
\mathcal{P}_2 =\{\{1, 4 ,5\},\{2, 3, 6\}\}
\end{align*}
\end{exmp}

For a given resolvable design  $(X, \mathcal{A})$ if |$X$| = $v$, |$\mathcal{A}$| = $b$ , block size is $k$ and number of parallel classes is $r$, then there are exactly $\frac{b}{r}$ blocks in each parallel class. Since the blocks in each parallel class are disjoint, therefore number of blocks in each parallel class is = $\frac{b}{r}$ = $\frac{v}{k}$.


\subsection{Cross Resolvable Design (CRD)\cite{KNRarXiv}}

\begin{defn}[\textbf{Cross Intersection Number}]
For any resolvable design $(X, \mathcal{A})$ with $r$ parallel classes, the $i^{th}$ cross intersection number, $\mu_{i}$ where $i \in \{2, 3, \dots,r\}$, is defined as the cardinality of intersection of $i$ blocks drawn from any $i$ distinct parallel classes, provided that, this value remains same ($\mu_i\neq 0$), for all possible choices of blocks.

For instance, in Example 1, $\mu_{2}$ = 1, as the intersection of any 2 blocks drawn from 2 distinct parallel classes is always at exactly one point. But we cannot define $\mu_{3}$ as the intersection of $3$ blocks drawn from 3 distinct parallel classes takes elements from the set \{0, 1\}.
\end{defn}
	
\begin{defn}[\textbf{Cross Resolvable Design}]
For any resolvable design $(X, \mathcal{A})$, if there exist at least one $i\in \{2,3,\dots,r\}$ such that the $i^{th}$ cross intersection number $\mu_{i}$ exists, then the resolvable design is said to be a Cross Resolvable Design (CRD). For a CRD the maximum value for $i$ for which $\mu_i$ exists is called the Cross Resolution Number (CRN) for that CRD. A CRD  with the CRN equal to $r$ is called a Maximal Cross Resolvable Design (MCRD).
\end{defn}

Note that the resolvable design in \textit{Example 2} is not a CRD  as $\mu_{2}$ does not exist.

\begin{exmp}
\label{exmp3}
For the resolvable design $(X, \mathcal{A})$ with $X = \;\{1, 2, 3, 4, 5, 6, 7, 8, 9\},$  and \\
$\mathcal{A} = \;\{\{1, 2, 3\},\{4, 5, 6\},\{7, 8, 9\}, \{1, 4, 7\},\{2, 5, 8\},\{3, 6, 9\}\},$ the parallel classes are $\mathcal{P}_1 = \;\{\{1, 2, 3\},\{4, 5, 6\},\{7, 8, 9\}\}$  and $ \mathcal{P}_2 =\; \{\{1, 4, 7\},\{2, 5, 8\},\{3, 6, 9\}\}$ and   $\mu_{2} = 1$.	
\end{exmp}

\begin{exmp}
\label{exmp4}
For the resolvable design $(X, \mathcal{A})$ with $X =\; \{1, 2, 3, 4, 5, 6, 7, 8\},$  and   \\ 
$\mathcal{A} =\; \{\{1, 2, 3, 4\},\{ 5, 6, 7, 8\},\{1, 2, 5, 6\}, \{3, 4, 7, 8\},\{1, 3, 5, 7\},\{2, 4, 6, 8\}\},$ the parallel classes are $\mathcal{P}_1 =\; \{\{1, 2, 3, 4\},\{5, 6, 7, 8\}\},$ $ \mathcal{P}_2 =\; \{\{1, 2, 5, 6\},\{3, 4, 7, 8\}\},$  and $ \mathcal{P}_3 =\; \{\{1, 3, 5, 7\},\{2, 4, 6, 8\}\}.$ In this case  $\mu_{2}$ = 2 and $\mu_{3}$ = 1.
\end{exmp}	

\begin{exmp}
\label{exmp5}
For the resolvable design $(X, \mathcal{A})$ with $X = \;\{1, 2, 3, 4, 5, 6, 7, 8, 9, 10, 11, 12\},$  and \\
$\mathcal{A} =\; \{\{1, 2, 3, 4, 5, 6\},\{ 7, 8, 9, 10, 11, 12\}, \{1, 2, 3, 7, 8, 9\},\{ 4, 5, 6, 10, 11, 12\}\}$ the parallel classes are
$\mathcal{P}_1 = \;\{\{1, 2, 3, 4, 5, 6\},\{ 7, 8, 9, 10, 11, 12\}\},$   and $ \mathcal{P}_2 =\; \{\{1, 2, 3, 7, 8, 9\},\{ 4, 5, 6, 10, 11, 12\}\}.$ We have $\mu_{2}$ = 3.
\end{exmp}

\begin{exmp}
\label{exmp6}
Consider the resolvable design $(X, \mathcal{A})$ with $ X = \;\{1, 2, 3, 4, 5, 6, 7, 8, 9\},$  and \\  $ \mathcal{A} =\{\{1, 2, 3\},\{4, 5, 6\},\{7, 8, 9\},\{1, 4, 7\},\{2, 5, 8\},\{3, 6, 9\}, \{1, 5, 9\},\{2, 6, 7\},\{3, 4, 8\},\{1, 6, 8\},$ $\{2, 4, 9\},\{3, 5, 7\}\}.$ The parallel classes are $\mathcal{P}_1 = \{\{1, 2, 3\},\{4, 5, 6\},\{7, 8, 9\}\},$\\ $\mathcal{P}_2 = \{\{1, 4, 7\},\{2, 5, 8\},\{3, 6, 9\}\},$ $\mathcal{P}_3 = \{\{1, 5, 9\},\{2, 6, 7\},\{3, 4, 8\}\},$  and \\ $\mathcal{P}_4 =\; \{\{1, 6, 8\},\{2, 4, 9\},\{3, 5, 7\}\}.$ Here  $\mu_{2} = 1$ and $\mu_{3}$, $\mu_{4}$ does not exist. 
\end{exmp}

From Example \ref{exmp6} one can see that $\mu_{r}$ need not always exist for a CRD.




\section{Coded Caching schemes from CRDs}
\label{Proposed scheme}

Given a cross resolvable design $(X, \mathcal{A})$ with $v$ points, $r$ parallel classes, $b$ blocks of size $k$ each, $b_r\stackrel{def}{=}\frac{b}{r}$ blocks in each parallel class, we choose some $z \in \{2,3,\dots,r\}$ such that $\mu_{z}$ exists. Let $\mathcal{A}_j$ denote the $j^{th}$ block in $\mathcal{A}$, assuming some ordering on the blocks of $\mathcal{A}$. We associate  a coded caching problem with $K = {r \choose z} (\binom{b_r}{t})^z$ number of users where, $t \in \{1,2,\dots,b_r\}$, $N$ files in server database, $b$  number of caches, $\frac{M}{N} = \frac{k}{v}$ fraction of each file at each cache and subpacketization level $v.$  A user is connected to distinct $tz$ caches such that these $tz$ caches correspond to distinct $t$ blocks from each of the distinct $z$ parallel classes. We denote the set of $K$ users  $\mathcal{K}$ as,
 $\mathcal{K} := \{U_{H} :\;|H| = tz\}$ where, $H$ is a $tz$ sized set containing cache indices from distinct parallel classes.


\subsection{Placement Phase}
\label{Placement} 

In the caching placement phase, we split each file $W_i,\;\forall i \in [N]$ into $v$ non-overlapping subfiles of equal size i.e.
$$W_i = (W_{i,k} : \forall k \in [v]),\; i = 1,2,\dots,N.$$
	
The placement is as follows. In the $j^{th}$ cache, the indices of the subfiles stored in $\mathcal{Z}_j$ is the $j^{th}$ block in the design. We assume symmetric batch prefetching i.e.,
$$\mathcal{Z}_j = \{{W}_{ik} : k \in \mathcal{A}_j\},\; \forall i \in \{1,2...,N\}, \forall j \in \{1,2...,b\}.$$
Therefore the total number of subfiles for each file in any cache is block size $k$ of the resolvable design i.e. $\frac{M}{N}=\frac{k}{v}$.  

Let $M'$ denote the size of the memory in units of files that a user has access to. We have

\begin{multline*}
\frac{M'}{N}= \sum_{i=1}^{tz} \frac{|\mathcal{A}_i|}{v}-\sum_{1\leq i_1<i_2\leq tz}^{tz} \frac{|\mathcal{A}_{i_1}\cap\mathcal{A}_{i_2}|}{v}+\dots +(-1)^{s+1}\\\sum_{1\leq i_1<\dots<i_s\leq tz}^{tz} \frac{|\mathcal{A}_{i_1}\cap\dots \cap\mathcal{A}_{i_s}|}{v}
+\dots+ (-1)^{tz+1}\frac{|\mathcal{A}_{1}\cap\dots\cap\mathcal{A}_{tz}|}{v}
\end{multline*}

where $\mathcal{A}_i,\;i\in[tz]$ are distinct $t$ blocks from each of distinct $z$ parallel classes. We get, 

\begin{multline*}
\frac{M'}{N}= zt\left(\frac{M}{N}\right) - (t^2)\binom{z}{2}\left( \frac{\mu_2}{v}\right) + \dots +(-1)^{s+1}(t^s)\binom{z}{s} \left(\frac{\mu_s}{v}\right) 
+\dots+ (-1)^{z+1}(t^z)\left(\frac{\mu_z}{v}\right),
\end{multline*}

which simplifies to
$$\frac{M'}{N}= \frac{ztM}{N}+\sum_{s=2}^z (-1)^{s+1}(t^s)\binom{z}{s}\left(\frac{\mu_s}{v}\right).$$



\subsection{Delivery Phase}
\label{Delivery}

For delivery, the users are arranged in lexicographical order of their indices $S$, establishing a one-to-one correspondence with the set $\{1, 2, \dots , K \}$.
We focus our attention to the case where user demands are distinct.
At the beginning of the delivery phase, each user requests one of the $N$ files. Let the demand vector be denoted by $\textbf{d} = (d_1,d_2,\dots,d_K)$.

The delivery steps are presented as an algorithm in {\bf Algorithm 1} the proof of correctness of which is given in the Appendix.
\begin{algorithm}[]
\caption{Delivery Algorithm}
\begin{algorithmic}[1]
\For {$u=1$ to $u=\binom{r}{z}$}
\State Choose any $z$ parallel classes out of $r$ parallel classes which is different 
\State from sets chosen before.
\State Let this set be $$\mathcal{P}_1 = \{C_{1,1},C_{1,2},\dots\dots,C_{1,b_r}\},$$
	   $$\mathcal{P}_2 = \{C_{2,1},C_{2,2},\dots\dots,C_{2,b_r}\},$$
	   $$\vdots$$
	   $$\mathcal{P}_z = \{C_{z,1},C_{z,2},\dots\dots,C_{z,b_r}\}.$$
	   \For{$v=1$ to $v=\binom{b_r}{t+1}^z$}
	   \State Choose $t+1$ of blocks from each of the parallel classes $\mathcal{P}_1,\mathcal{P}_2,\dots,\mathcal{P}_z$. 
	   \State This set of $(t+1)z$ blocks must be different from the ones chosen before. 
	   \State Let the chosen set be  
	   \State $C_{1,i_{1_0}},C_{1,i_{1_1}},\dots,C_{1,i_{1_t}}$, $C_{2,i_{2_0}},C_{2,i_{2_1}},\dots,C_{2,i_{2_t}}$,  $\dots,C_{z,i_{z_0}},C_{z,i_{z_1}},\dots,C_{z,i_{z_t}}$,
	   \State where $i_{s_k} \in [1,b_r] \text{ and } i_{s_k} \neq i_{s_{k'}},\;\forall k,\;k' \in [0,t] \text{ and } \forall s \in [1,z]$.
	   \State There are $(t+1)^z$ users corresponding to the $(t+1)z$ blocks chosen above. 
	   \State Denoting this set of user indices by $\mathcal{X}$.
	   \State \textbf{Calculate:} Calling the user connected to the set of caches
	   \State $C_{1,a_{1_1}},C_{1,a_{1_2}},\dots,C_{1,a_{1_t}}$, $C_{2,a_{2_1}},C_{2,a_{2_2}},\dots,C_{2,a_{2_t}}$, $\dots,C_{z,a_{z_1}},C_{z,a_{z_2}},\dots,C_{z,a_{z_t}}$, 
	   \State where $a_{s_k}\in \{i_{s_0},i_{s_1},\dots,i_{s_t}\} \text{ and } a_{s_k}\neq a_{s_{k'}},\;\forall k,\;k' \in [1,t] \text{ and }\forall s = [1,z]$,
	   \State to be $m^{th}$ user calculate the set $f_m$ as $f_m = C_{1,e_1}\cap C_{2,e_2}\cap\dots\cap C_{z,e_{z}}$
	   \State where $e_s = \{i_{s_0},i_{s_1},\dots,i_{s_t}\}\setminus\{a_{s_1},a_{s_2},\dots,a_{s_t}\},\;\forall s\in [1,z]$. We have 
	   \State $|f_m|=\mu_{z}$. Let $f_m := \{y_{m,1},y_{m,2},\dots,y_{m,\mu_{z}}\}$.
	   \State Calculate $f_m$ as above for all the $(t+1)^z$ users in $\mathcal{X}$.
	   \State \textbf{Transmit:} Now do the following $\mu_z$ transmissions
	   \State $$\underset {m \in \mathcal{X}} \oplus W_{d_{m},y_{m,s}},\;\forall s \in [\mu_{z}]$$
	   \State Note that there are $\mu_{z}$ transmissions for $\mathcal{X}$.
	   \EndFor
\EndFor
\end{algorithmic}
\end{algorithm}
	
\begin{thm}
	For $N$ files and $K$ users each with access to $zt$ caches of size $M$ in the considered caching system, if $N\geq K$ and for the distinct demands by the users, the proposed scheme achieves the rate $R$ given by
	$$R = \frac{\mu_z\binom{b_r}{t + 1}^z\binom{r}{z}}{v}$$
\end{thm}

\begin{IEEEproof}
	The first \textbf{for} loop of the delivery algorithm runs $\binom{r}{z}$ times. The second \textbf{for} loop of the delivery algorithm runs $\binom{b_r}{t+1}^z$ times. The transmit step of the delivery algorithm runs $\mu_z$ times.
	So we see that totally there are $\frac{\mu_z\binom{b_r}{t + 1}^z\binom{r}{z}}{v}$ transmissions and subpacketization level is $v$. 
\end{IEEEproof}

\begin{lem}
	\label{lemgain}
	The number of users benefited in each transmission, known in the literature as coding gain and denoted by $g,$ is given by $ g = (t+1)^{z}.$
\end{lem}

\begin{IEEEproof}
	From second \textbf{for} of the delivery algorithm it can be observed that, there are totally $(t+1)^{z}$ users benefited from a transmission. So the coding gain by definition is $(t+1)^{z}$.
\end{IEEEproof}

\begin{rem}
	In the proposed scheme, when $t = 1$, the scheme in \cite{KNRarXiv} is obtained. 
\end{rem}


\section{Performance Comparison}


For performance comparison, we use the CRDs from affine resolvable designs obtained through affine geometry.
Such CRDs exists for all $q$ and $m,$ where $q$ is a prime or a power of a prime number and $m\geq2$. The number of points $v=q^m$, the number of points in a block $k=q^{m-1}$, the number of blocks $b=\frac{q(q^m-1)}{q-1}$, the number of parallel classes $r=\frac{q^m-1}{q-1}.$ It is known that any two blocks drawn from different parallel classes always intersect at exactly $\frac{k^2}{v}$ points \cite{Stinson} i.e., $z=2$ and $\mu_z=q^{m-2}.$

\begin{figure}[]
	\begin{center}
		\includegraphics[width=12cm,height=8.5cm]{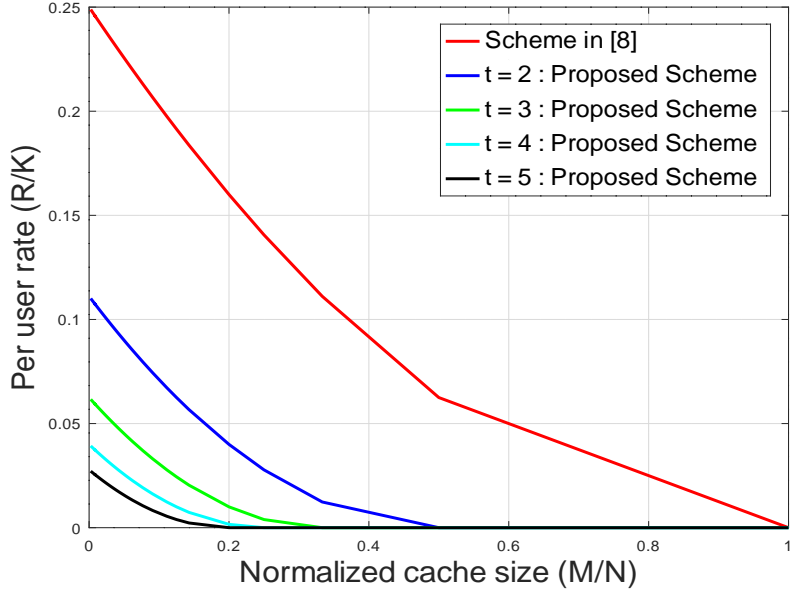}
		\caption {Comparison of the scheme in \cite{KNRarXiv} and proposed scheme for the class of cross resolvable design derived from affine resolvable BIBD's for the case $z=2$,  where q is a prime or prime power and $m \geq 2$.}
		\label{Extended_CRD_vs_CRD}
	\end{center}
\end{figure}



\subsubsection{Comparison with the scheme in \cite{KNRarXiv}}
For the multi-access coded caching scheme from CRDs with parameters $q$ and $m,$ we have, $K=(b_r)^z\binom{r}{z}=\frac{q(q^m-1)(q^{m-1}-1)\binom{q}{t}^2}{2(q-1)^2}$ users, $b=\frac{q(q^m-1)}{q-1}$ caches each having a cache size of $\frac{M}{N}=\frac{k}{v}=\frac{1}{q}$ and $\frac{M'}{N} = \frac{(2qt-t^2)}{q^2}.$

Fig \ref{Extended_CRD_vs_CRD} shows the variation of per user rate versus normalized cache size $\frac{M}{N}$. It can be seen that per-user-rate decreases as $t$ increases.

\begin{figure}[]
	\begin{center}
		\includegraphics[width=12cm,height=9cm]{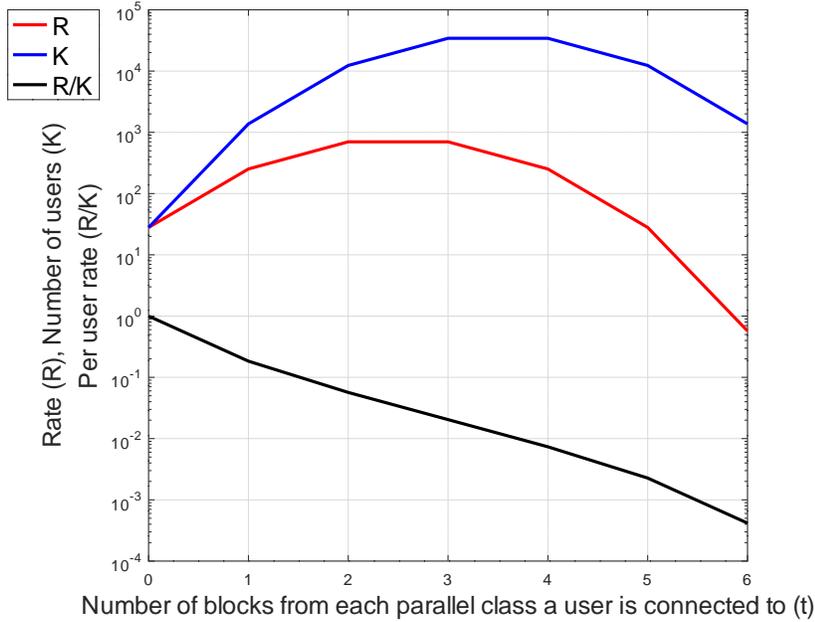}
		\caption {Comparison of Rate $(R)$, Number of users $(K)$ and Per user rate $\frac{R}{K}$ with respect to $t$ for CRDs with parameters $v=49$, $b=56$, $k=7$, $r=8$, $z=2$ and $\mu_2 = 1$}
		\label{Extended_CRD}
	\end{center}
\end{figure}

Fig \ref{Extended_CRD}, shows how $K$, $R$ and $\frac{R}{K}$ varies with respect to $t$. The case when $t = 1$, corresponds to the scheme in \cite{KNRarXiv}. It is seen that the per user rate decreases as $t$ increases. The number of users increase as $t$ increases and then reaches a maximum after which it starts decreasing due to the binomial coefficient in the expression for $K$. Similar behavior can be seen for $R$ also.

\begin{figure}[]
	\begin{center}
		\includegraphics[width=12cm,height=9cm]{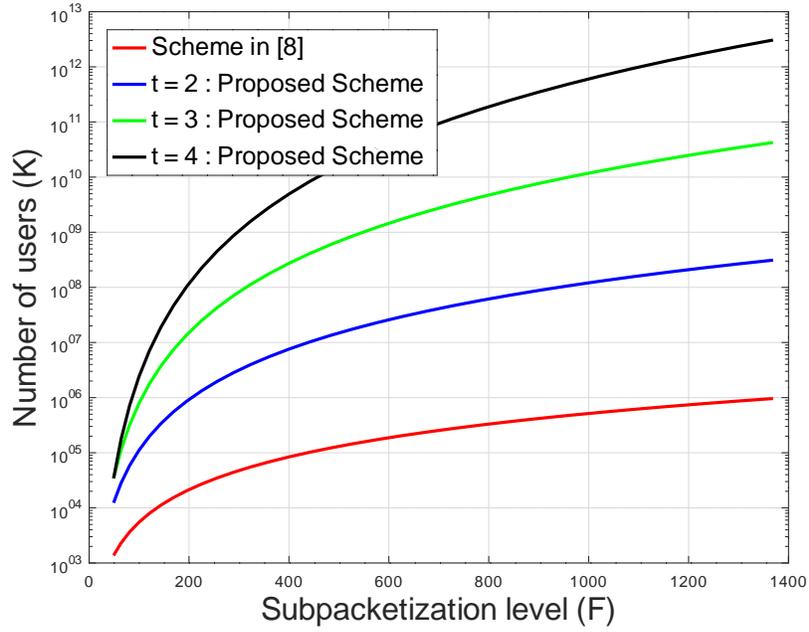}
		\caption {Comparison of subpacketization levels of the scheme  in \cite{KNRarXiv} and proposed scheme, $t\in\{2,3,4\}$ for the class of resolvable designs derived from affine resolvable BIBD's , where q is a prime or prime power and $m=2$.}
		\label{Extended_CRD_vs_CRD_F}
	\end{center}
\end{figure}

Fig. \ref{Extended_CRD_vs_CRD_F} depicts subpacketization $(F)$ versus number of users for different values of $t$. It can be seen that for the same subpacketization levels, the number of users increases as $t$ increases .

\subsubsection{Comparison with the MaN scheme}
Since $\frac{bM}{N}=\frac{b}{q} = \frac{(q^m-1)}{q-1}$ is a integer, 
there exist a coded caching system with users equipped with dedicated caches with
 $\frac{q(q^m-1)}{q-1}$ users and $\frac{M}{N} = \frac{1}{q}$.
 Table \ref{tab1} shows the comparison of the MaN scheme and proposed scheme.

\begin{table}
\caption{Comparison of MaN scheme and proposed scheme for $q$ being a prime or prime power and $m \geq 2$}
\begin{center}
\renewcommand{\arraystretch}{2}
\begin{tabular}{|c|c|c|}
\hline
\textbf{Parameters} &\textbf{MaN Scheme} & \textbf{Proposed Scheme}\\\hline\hline
\makecell{Number of Caches} & $\dfrac{q(q^m-1)}{q-1}$ & $\dfrac{q(q^m-1)}{q-1}$\\[.2cm]\hline
 Number of caches a user has access to & $1$ & $2t$ \\\hline
\makecell{Fraction of each file at each cache $\left(\frac{M}{N}\right)$} & $\dfrac{1}{q}$ & $\dfrac{1}{q}$\\\hline
\makecell{Number of Users $(K)$} & $\dfrac{q(q^m-1)}{q-1}$ &$\dfrac{q(q^m-1)(q^{m-1}-1)\binom{q}{t}^2}{2(q-1)^2}$\\[.2cm]\hline
\makecell{Subpackelization level $(F)$} & $\dbinom{q(q^m-1)/q-1}{(q^m-1)/q-1}$ & $q^m$\\[.2cm]\hline
Rate $(R)$ & $\dfrac{(q^m-1)(q-1)}{q^m+q-2}$ & $\dfrac{(q^m-1)(q^{m-1}-1)\binom{q}{t+1}^2}{2q(q-1)^2}$\\[.2cm]\hline
\makecell{Rate per\ user $\left(\frac{R}{K}\right)$}&$\dfrac{(q-1)^2}{q(q^m+q-2)}$&$\left(\dfrac{\binom{q}{t+1}}{q\binom{q}{t}}\right)^2$\\[.2cm]\hline
Gain $(g)$ & $\dfrac{q^m-1}{q-1}+1$ & $(t+1)^2$\\[.2cm]\hline
\end{tabular}
\end{center}
\label{tab1}
\end{table}

\begin{figure}[]
	\begin{center}
		\includegraphics[width=12cm,height=8cm]{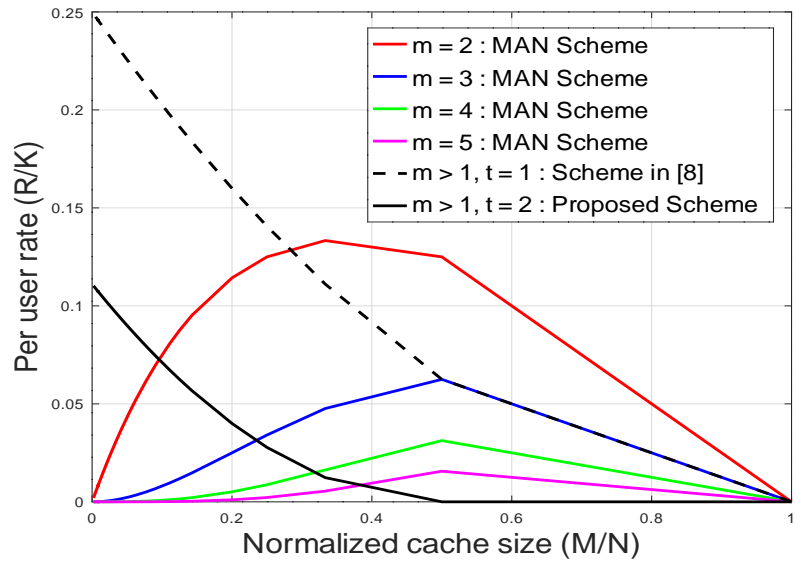}
		\caption {Comparison between MaN and Proposed scheme for the class of cross resolvable design derived from affine resolvable BIBD's for the case $z=2$,  where q is a prime or prime power and $m \geq 2$.}
		\label{Extended_CRD_vs_MAN}
	\end{center}
\end{figure}
Fig. \ref{Extended_CRD_vs_MAN} shows the variation of per user rate $\frac{R}{K}$ versus the fraction of each file stored at each cache $\frac{M}{N}$. Since $\frac{M}{N} = \frac{1}{q}$ and $\frac{R}{K} = \frac{(q-1)^2}{(q)(q^m+q-2)}$ in the case of the MaN scheme, and $\frac{R}{K} = \left(\dfrac{\binom{q}{t+1}}{q\binom{q}{t}}\right)^2$ in case of proposed scheme is a function of $q$, we plot $\frac{R}{K}$ vs $\frac{M}{N}$ keeping $m$ constant for different values of $q$, where $q$ is a prime or prime power. 

\section{EXAMPLES}
\begin{exmp}
\label{exmp7}
Consider the resolvable design $(X, \mathcal{A})$ with   
\begin{align*}
    X = \;& \;\{1,2,3,4,5,6,7,8,9,10,11,12,13,14,15,16,17,18,19,20,21,22,23,24,25,26,27\}\\
    \mathcal{A} =\; & \;\{\{1, 4, 7, 10, 13, 16, 19, 22, 25\},\{2, 5, 8, 11, 14, 17, 20, 23, 26\},\{3, 6, 9, 12, 15, 18, 21, 24, 27\},\\ & \;\;\;\{1,2,3,10,11,12,19,20,21\},\{4,5,6,13,14,15,22,23,24\},\{7,8,9,16,17,18,25,26,27\},\\ & \;\;\;\{1,2,3,4,5,6,7,8,9\},\{10,11,12,13,14,15,16,17,18\},\{19,20,21,22,23,24,25,26,27\}\}
\end{align*}
The parallel classes are \\
$\mathcal{P}_1 = \{\underbrace{\{1,4,7,10,13,16,19,22,25\}}_{C_{1,1}},\underbrace{\{2,5,8,11,14,17,20,23,26\}}_{C_{1,2}},\underbrace{\{3,6,9,12,15,18,21,24,27\}}_{C_{1,3}}\}$\\\\
$\mathcal{P}_2 = \{\underbrace{\{1,2,3,10,11,12,19,20,21\}}_{C_{2,1}},\underbrace{\{4,5,6,13,14,15,22,23,24\}}_{C_{2,2}},\underbrace{\{7,8,9,16,17,18,25,26,27\}}_{C_{2,3}}\}$\\\\
$\mathcal{P}_3 = \{\underbrace{\{1,2,3,4,5,6,7,8,9\}}_{C_{3,1}},\underbrace{\{10,11,12,13,14,15,16,17,18\}}_{C_{3,2}},\underbrace{\{19,20,21,22,23,24,25,26,27\}}_{C_{3,3}}\}$\\\\
Here $v = 27$, $b = 9$, $k = 9$, $r = 3$,  $\mu_{2} = 3$ and $\mu_{3}$ = 1. Let $z = 2$ and $t = 2$ therefore, fraction of each file at each cache $\frac{M}{N} = \frac{k}{v} = \frac{1}{3}$ and number of users $K = {r \choose z} \left(\binom{b_r}{t}\right)^z = {3 \choose 2} (\binom{3}{2})^2 = 27$.
Now consider the user $U_{\{C_{1,1},C_{1,2},C_{2,1},C_{2,2}\}}$ connected to caches $C_{1,1},C_{1,2},C_{2,1} \text{ and } C_{2,2}$,  it can be verified that user $U_{\{C_{1,1},C_{1,2},C_{2,1},C_{2,2}\}}$ has access to subfiles indices $\mathcal{Z}_{U_{\{C_{1,1},C_{1,2},C_{2,1},C_{2,2}\}}} = X\setminus\{9,18,27\}$ such that $|\mathcal{Z}_{U_{\{C_{1,1},C_{1,2},C_{2,1},C_{2,2}\}}}| = 24$, which can also be verified from the equation
$$\frac{M'}{N}  = \frac{ztM}{N}+\sum_{s=2}^z (-1)^{s+1}(t^s)\binom{z}{s}\left(\frac{\mu_s}{v}\right)
     = \frac{4}{3}+ (-1)^{3}(2^2)\binom{2}{2}\left(\frac{3}{27}\right) = \frac{24}{27}
$$
\textit{Delivery Phase:} In the first for loop of \textbf{Algorithm 1} we select $z=2$ parallel classes out of $r=3$. Let the chosen parallel classes be $\mathcal{P}_1$ and $\mathcal{P}_2$. In the second for loop of \textbf{Algorithm 1} we select $t+1=3$ blocks from each of the parallel class, $\mathcal{P}_1$ and $\mathcal{P}_2$. Let the chosen blocks are $C_{1,1},C_{1,2},C_{1,3},C_{2,1},C_{2,2}$ and $C_{2,3}$. Corresponding to these $(t+1)z = 6$ blocks chosen above, we have $(t+1)^z=9$ users given by the set
\begin{align*}
    \mathcal{X} = \;&\; \{\underbrace{U_{\{C_{1,1},C_{1,2},C_{2,1},C_{2,2}\}}}_{U_{1}},\;\underbrace{U_{\{C_{1,1},C_{1,2},C_{2,1},C_{2,3}\}}}_{U_{2}},\;\underbrace{U_{\{C_{1,1},C_{1,2},C_{2,2},C_{2,3}\}}}_{U_{3}},\\
    \;&\; \{\underbrace{U_{\{C_{1,1},C_{1,3},C_{2,1},C_{2,2}\}}}_{U_{4}},\;\underbrace{U_{\{C_{1,1},C_{1,3},C_{2,1},C_{2,3}\}}}_{U_{5}},\;\underbrace{U_{\{C_{1,1},C_{1,3},C_{2,2},C_{2,3}\}}}_{U_{6}},\\
    \;&\; \{\underbrace{U_{\{C_{1,2},C_{1,3},C_{2,1},C_{2,2}\}}}_{U_{7}},\;\underbrace{U_{\{C_{1,2},C_{1,3},C_{2,1},C_{2,3}\}}}_{U_{8}},\;\underbrace{U_{\{C_{1,2},C_{1,3},C_{2,2},C_{2,3}\}}}_{U_{9}}\}
\end{align*}
Subfiles indices accessible to $m^{th}$ user in the set $\mathcal{X}$ are given by
$$\mathcal{Z}_{U_1} = X\setminus\{9,18,27\},\;\; \mathcal{Z}_{U_2} = X\setminus\{6,15,24\},\;\; \mathcal{Z}_{U_3} = X\setminus\{3,12,21\},$$
$$\mathcal{Z}_{U_4} = X\setminus\{8,17,26\},\;\; \mathcal{Z}_{U_5} = X\setminus\{5,14,23\},\;\; \mathcal{Z}_{U_6} = X\setminus\{2,11,20\},$$
$$\mathcal{Z}_{U_7} = X\setminus\{7,16,25\},\;\; \mathcal{Z}_{U_8} = X\setminus\{4,13,22\},\;\; \mathcal{Z}_{U_9} = X\setminus\{1,10,19\}$$
Now we calculate the set $f_m,\;m\in\mathcal{X}$. For the user $U_{\{C_{1,1},C_{1,2},C_{2,1},C_{2,2}\}} \equiv U_1$ set $f_{U_{1}}$ is calculated as $f_{U_{1}} = \{C_{1,3}\cap C_{2,3}\} = \{9,18,27\}$. Also $|f_{U_1}| = 3$. Similarly for all the users in set $\mathcal{X}$, we have
$$f_{U_2}=\{6,15,24\},f_{U_3}=\{3,12,21\},f_{U_4}=\{8,17,26\},f_{U_5}=\{5,14,23\},$$
$$f_{U_6}=\{2,11,20\},f_{U_7}=\{7,16,25\},f_{U_8}=\{4,13,22\},f_{U_9}=\{1,10,19\}$$
It can checked that subfiles indices in the set $f_{U_{m}},\;m\;\in\;\mathcal{X}$ are not accessible to the user $U_m$ i.e $f_{U_{m}}\cap \mathcal{Z}_{U_{m}} = \emptyset, \;\; \forall m \; \in \; \mathcal{X}$. It can also be verified that 
$f_m \cap \mathcal{Z}_{m':\;\forall m'\;\in\;\mathcal{X}\setminus m} = f_m, \;\; \forall m \; \in \; \mathcal{X}$. Let the demand of the $m^{th}$ user in $\mathcal{X}$ is given by $d_{U_m} \equiv\; d_m, \;\forall m \in \mathcal{X}$.
Transmissions corresponding to the set $\mathcal{X}$ are:
$$W_{d_1,9}\oplus W_{d_2,6}\oplus W_{d_3,3}\oplus W_{d_4,8}\oplus W_{d_5,5}\oplus W_{d_6,2}\oplus W_{d_7,7}\oplus W_{d_8,4}\oplus W_{d_9,1},$$
$$W_{d_1,18}\oplus W_{d_2,15}\oplus W_{d_3,12}\oplus W_{d_4,17}\oplus W_{d_5,14}\oplus W_{d_6,11}\oplus W_{d_7,16}\oplus W_{d_8,13}\oplus W_{d_9,10} \;\text{ and }$$
$$W_{d_1,27}\oplus W_{d_2,24}\oplus W_{d_3,21}\oplus W_{d_4,26}\oplus W_{d_5,23}\oplus W_{d_6,20}\oplus W_{d_7,25}\oplus W_{d_8,22}\oplus W_{d_9,19}$$

From the above $|f_m| = 3$ transmissions it can be easily seen that all the users in set the $\mathcal{X}$ will receive all the subfiles of the demanded file.
\end{exmp}

\begin{exmp}
\label{exmp8}
Consider the resolvable design $(X, \mathcal{A})$ with   
\begin{align*}
    X = \;& \;\{1,2,3,4,5,6,7,8,9,10,11,12,13,14,15,16\}\\
    \mathcal{A} =\; & \;\{\{1,5,9,13\},\{2,6,10,14\},\{3,7,11,15\},\{4,8,12,16\},\{1,2,3,4\},\{5,6,7,8\},\\ &  \;\;\;\{9,10,11,12\},\{13,14,15,16\},\{1,6,11,16\},\{2,7,12,13\},\{3,8,9,14\},\{4,5,10,15\}\}
\end{align*}
The parallel classes are
$$\mathcal{P}_1 = \{\underbrace{\{1,2,3,4\}}_{C_{1,1}},\underbrace{\{5,6,7,8\}}_{C_{1,2}},\underbrace{\{9,10,11,12\}}_{C_{1,3}},\underbrace{\{13,14,15,16\}}_{C_{1,4}}\}$$
$$\mathcal{P}_2 = \{\underbrace{\{1,5,9,13\}}_{C_{2,1}},\underbrace{\{2,6,10,14\}}_{C_{2,2}},\underbrace{\{3,7,11,15\}}_{C_{2,3}},\underbrace{\{4,8,12,16\}}_{C_{2,4}}\}$$
$$\mathcal{P}_3 = \{\underbrace{\{1,6,11,16\}}_{C_{3,1}},\underbrace{\{2,7,12,13\}}_{C_{3,2}},\underbrace{\{3,8,9,14\}}_{C_{3,3}},\underbrace{\{4,5,10,15\}}_{C_{3,4}}\}$$
Here $v = 16$, $b = 12$, $k = 4$, $r = 3$,  $\mu_{2} = 1$ and $\mu_{3}$ does not exist. Let $z = 2$ and $t = 2$ therefore, fraction of each file at each cache $\frac{M}{N} = \frac{k}{v} = \frac{1}{4}$ and number of users $K = {r \choose z} \left(\binom{b_r}{t}\right)^z = {3 \choose 2} (\binom{4}{2})^2 = 108$.
Now consider the user $U_{\{C_{1,1},C_{1,2},C_{2,1},C_{2,2}\}}$ connected to caches $C_{1,1},C_{1,2},C_{2,1} \text{ and } C_{2,2}$,  it can be verified that user $U_{\{C_{1,1},C_{1,2},C_{2,1},C_{2,2}\}}$ has access to subfiles indices $\mathcal{Z}_{U_{\{C_{1,1},C_{1,2},C_{2,1},C_{2,2}\}}} = X\setminus\{11,12,15,16\}$ such that $|\mathcal{Z}_{U_{\{C_{1,1},C_{1,2},C_{2,1},C_{2,2}\}}}| = 12$, which can also be verified from the equation
$$\frac{M'}{N}  = \frac{ztM}{N}+\sum_{s=2}^z (-1)^{s+1}(t^s)\binom{z}{s}\left(\frac{\mu_s}{v}\right)
     = 1+ (-1)^{3}(2^2)\binom{2}{2}\left(\frac{1}{16}\right) = \frac{3}{4}
$$

\textit{Delivery Phase:} In the first for loop of \textbf{Algorithm 1} we select $z=2$ parallel classes out of $r=3$. Let the chosen parallel classes be $\mathcal{P}_1$ and $\mathcal{P}_2$. In the second for loop of \textbf{Algorithm 1} we select $t+1=3$ blocks from each of the parallel class. Let the chosen blocks are $C_{1,1},C_{1,2},C_{1,3},C_{2,1},C_{2,2}$ and $C_{2,3}$. Corresponding to these $(t+1)z = 6$ blocks we have $(t+1)^z=9$ users given by the set
\begin{align*}
    \mathcal{X} = \;&\; \{\underbrace{U_{\{C_{1,1},C_{1,2},C_{2,1},C_{2,2}\}}}_{U_{1}},\;\underbrace{U_{\{C_{1,1},C_{1,2},C_{2,1},C_{2,3}\}}}_{U_{2}},\;\underbrace{U_{\{C_{1,1},C_{1,2},C_{2,2},C_{2,3}\}}}_{U_{3}},\\
    \;&\; \{\underbrace{U_{\{C_{1,1},C_{1,3},C_{2,1},C_{2,2}\}}}_{U_{4}},\;\underbrace{U_{\{C_{1,1},C_{1,3},C_{2,1},C_{2,3}\}}}_{U_{5}},\;\underbrace{U_{\{C_{1,1},C_{1,3},C_{2,2},C_{2,3}\}}}_{U_{6}},\\
    \;&\; \{\underbrace{U_{\{C_{1,2},C_{1,3},C_{2,1},C_{2,2}\}}}_{U_{7}},\;\underbrace{U_{\{C_{1,2},C_{1,3},C_{2,1},C_{2,3}\}}}_{U_{8}},\;\underbrace{U_{\{C_{1,2},C_{1,3},C_{2,2},C_{2,3}\}}}_{U_{9}}\}
\end{align*}
Subfiles indices accessible to $m^{th}$ user in the set $\mathcal{X}$ are given by
$$\mathcal{Z}_{U_1} = X\setminus\{11,12,15,16\},\;\; \mathcal{Z}_{U_2} = X\setminus\{10,12,14,16\},\;\; \mathcal{Z}_{U_3} = X\setminus\{9,12,13,16\},$$
$$\mathcal{Z}_{U_4} = X\setminus\{7,8,15,16\},\;\; \mathcal{Z}_{U_5} = X\setminus\{6,8,14,16\},\;\; \mathcal{Z}_{U_6} = X\setminus\{5,8,13,16\},$$
$$\mathcal{Z}_{U_7} = X\setminus\{3,4,15,16\},\;\; \mathcal{Z}_{U_8} = X\setminus\{2,4,14,16\},\;\; \mathcal{Z}_{U_9} = X\setminus\{1,4,13,16\}$$
Now we calculate the set $f_m,\;m\in\mathcal{X}$. For the user $U_{\{C_{1,1},C_{1,2},C_{2,1},C_{2,2}\}} \equiv U_1$ set $f_{U_{1}}$ is calculated as $f_{U_{1}} = \{C_{1,3}\cap C_{2,3}\} = \{11\}$. Also $|f_{U_1}| = 1$. Similarly for all the users in set $\mathcal{X}$, we have
$$f_{U_2}=\{10\},f_{U_3}=\{9\},f_{U_4}=\{7\},f_{U_5}=\{6\},$$
$$f_{U_6}=\{5\},f_{U_7}=\{3\},f_{U_8}=\{2\},f_{U_9}=\{1\}$$
It can checked that subfiles indices in the set $f_{U_{m}},\; m \in \mathcal{X}$ are not accessible to the user $U_m$ i.e $f_{U_{m}}\cap \mathcal{Z}_{U_{m}} = \emptyset,\;\; \forall m  \in  \mathcal{X}$. It can also be verified that 
$f_m \cap \mathcal{Z}_{m':\;\forall m'\;\in\;\mathcal{X}\setminus m} = f_m, \;\; \forall m  \in  \mathcal{X}$. Let the demand of the $m^{th}$ user in $\mathcal{X}$ is given by $d_{U_m} \equiv\; d_m, \;\forall m \in \mathcal{X}$.
Transmission corresponding to the set $\mathcal{X}$ are:
$$W_{d_1,11}\oplus W_{d_2,10}\oplus W_{d_3,9}\oplus W_{d_4,7}\oplus W_{d_5,6}\oplus W_{d_6,5}\oplus W_{d_7,3}\oplus W_{d_8,2}\oplus W_{d_9,1},$$
From the above $|f_m| = 1$ transmissions it can be easily seen that all the users in set the $\mathcal{X}$ will not receive all the subfiles of the demanded file. These users will receive other subfiles from the other $(t+1)z = 6$ combination of caches from the parallel class $\mathcal{P}_1$ and $\mathcal{P}_2$. 
\end{exmp}

\section{CONCLUSION}
In \cite{KNRarXiv} multi-access coded caching scheme is presented by taking one block from each of the parallel classes of the CRD used for the scheme. In this paper it is shown that by allowing more than one block from each of the parallel classes lower per-user-rate can be obtained without increasing the subpacketization level. 

\section*{Acknowledgment}
This work was supported partly by the Science and Engineering Research Board (SERB) of Department of Science and Technology (DST), Government of India, through J.C. Bose National Fellowship to B. Sundar Rajan.

\appendix

\begin{center}
\bf{Proof of Correctness of algorithm :}
\end{center}

\begin{lem} Let ${Y_{m}},\;m \in \mathcal{K}$ be the set of indices of subfiles accessible to the $m^{th}$ user. It is easily seen that $|Y_m|=\frac{vM'}{N}.$ Consider the transmission corresponding to the set $\mathcal{X}$ and a user $m\in \mathcal{X}$ as in {\bf Algorithm 1}. Then the following equality holds.
	\begin{equation}
	\label{fmequality}
	f_{m} = \underset{t \in \mathcal{X}\setminus m} \cap  {Y_{t}},\;\forall m \in \mathcal{X}
	\end{equation}
\end{lem}

\begin{IEEEproof} In \textbf{Algorithm 1} consider the combination of $(t+1)z$ blocks (caches) $C_{1,i_{1_0}},C_{1,i_{1_1}},\dots,C_{1,i_{1_t}}$, $C_{2,i_{2_0}},C_{2,i_{2_1}},\dots,C_{2,i_{2_t}}$, $\dots,C_{z,i_{z_0}},C_{z,i_{z_1}},\dots,C_{z,i_{z_t}}$,
	where $i_{s_k} \in [1,b_r]$  and  $i_{s_k} \neq i_{s_{k'}},\;\forall k,\;k' \in [0,t]$ and $\forall s \in [1,z]$ and the $m^{th}$ user which has access to $tz$ blocks  $C_{1,a_{1_1}},C_{1,a_{1_2}},\dots,C_{1,a_{1_t}}$, $C_{2,a_{2_1}},C_{2,a_{2_2}},\dots,C_{2,a_{2_t}}$, $\dots,C_{z,a_{z_1}}C_{z,a_{z_2}},\dots, C_{z,a_{z_t}}$, where $a_{s_k}\in \{i_{s_0},i_{s_1},\dots,i_{s_t}\}$ and  $a_{s_k}\neq a_{s_{k'}},\;\forall k,\;k' \in [1,t]$ and $\forall s = [1,z]$.
	
	$$f_m = \bigcap_{\substack{
			l_{s_k}\;=\;\{i_{s_0},i_{s_1},\dots,i_{s_t}\},\\l_{s_k}\;\neq\;l_{s_{k'}},\\\forall k,\;k'\in[1,t]\text{ and }\forall s \;\in\; [1,z] \\ (l_{1_1},l_{1_2},\dots,l_{1_t},\dots,l_{z_1},l_{z_2},\dots,l_{z_t})\;\neq\\(a_{1_1},a_{1_2},\dots,a_{1_t},\dots,a_{z_1},a_{z_2},\dots,a_{z_t})}}\{C_{1,l_{1_1}}\cup C_{1,l_{1_2}}\cup\dots\cup C_{z,l_{1_t}}\cup\dots \cup C_{z,l_{z_1}}\cup C_{z,l_{z_2}}\cup\dots\cup C_{z,l_{z_t}}\}$$
	
	\begin{multline*}
	=\left\{\bigcap_{\substack{
			l_{s_k}\;=\;\{i_{s_0},i_{s_1},\dots,i_{s_t}\},\\l_{s_k}\;\neq\;l_{s_{k'}},\\\forall k,\;k'\in[1,t],\;\forall s \;\in\; [1,z] \\ (l_{1_1},l_{1_2},\dots,l_{1_t})\;\neq\\(a_{1_1},a_{1_2},\dots,a_{1_t})}}\{C_{1,l_{1_1}}\cup C_{1,l_{1_2}}\cup \dots\cup C_{1,l_{1_t}}\cup\dots\cup C_{z,l_{z_t}}\}\right\}\bigcap\\
	\left\{\bigcap_{\substack{
			l_{s_k}\;=\;\{i_{s_0},i_{s_1},\dots,i_{s_t}\},\\l_{s_k}\;\neq\;l_{s_{k'}},\\\forall k,\;k'\in[1,t]\text{ and }\forall s \;\in\; [2,z] \\ (l_{2_1},l_{2_2},\dots,l_{2_t},\dots,l_{z_1},l_{z_2},\dots,l_{z_t})\;\neq\\(a_{2_1},a_{2_2},\dots,a_{2_t},\dots,a_{z_1},a_{z_2},\dots,a_{z_t})}}\{C_{1,a_{1_1}}\cup C_{1,a_{12}}\cup \dots\cup C_{1,a_{1t}}\cup\dots\cup C_{z,l_{z_t}}\}\right\}\\   
	\end{multline*}
	
	\begin{multline*}
	f_m=\left\{\bigcap_{\substack{
			l_{1_k}\;=\;\{i_{1_0},i_{1_1},\dots,i_{1_t}\},\\l_{1_k}\;\neq\;l_{1_{k'}},\\\forall k,\;k'\in[1,t], \\ (l_{1_1},l_{1_2},\dots,l_{1_t})\;\neq\\(a_{1_1},a_{1_2},\dots,a_{1_t})}}\{C_{1,l_{1_1}}\cup C_{1,l_{1_2}}\cup\dots\cup C_{1,l_{1_t}}\}\right\}\bigcap\\
	\left\{\bigcap_{\substack{
			l_{s_k}\;=\;\{i_{s_0},i_{s_1},\dots,i_{s_t}\},\\l_{s_k}\;\neq\;l_{s_{k'}},\\\forall k,\;k'\in[1,t]\text{ and }\forall s \;\in\; [2,z] \\ (l_{2_1},l_{2_2},\dots,l_{2_t},\dots,l_{z_1},l_{z_2},\dots,l_{z_t})\;\neq\\(a_{2_1},a_{2_2},\dots,a_{2_t},\dots,a_{z_1},a_{z_2},\dots,a_{z_t})}}\{C_{1,a_{1_1}}\cup C_{1,a_{1_2}}\cup \dots\cup C_{1,a_{1_t}}\cup\dots\cup C_{z,l_{z_t}}\}\right\} 
	\end{multline*}

	$$f_m=C_{1,e_1}\bigcap
	\left\{\bigcap_{\substack{
			l_{s_k}\;=\;\{i_{s_0},i_{s_1},\dots,i_{s_t}\},\\l_{s_k}\;\neq\;l_{s_{k'}},\\\forall k,\;k'\in[1,t]\text{ and }\forall s \;\in\; [2,z] \\ (l_{2_1},l_{2_2},\dots,l_{2_t},\dots,l_{z_1},l_{z_2},\dots,l_{z_t})\;\neq\\(a_{2_1},a_{2_2},\dots,a_{2_t},\dots,a_{z_1},a_{z_2},\dots,a_{z_t})}}\{C_{1,a_{1_1}}\cup C_{1,a_{1_2}}\cup \dots\cup C_{1,a_{1_t}}\cup\dots\cup C_{z,l_{z_t}}\}\right\}$$

	$$f_m=C_{1,e_1}\bigcap\left\{\{C_{1,a_{1_1}}\cup C_{1,a_{1_2}}\cup \dots\cup C_{1,a_{1_t}}\}\cup
	\left\{\bigcap_{\substack{l_{s_k}\;=\;\{i_{s_0},i_{s_1},\dots,i_{s_t}\},\\l_{s_k}\;\neq\;l_{s_{k'}},\\\forall k,\;k'\in[1,t]\text{ and }\forall s \;\in\; [2,z] \\ (l_{2_1},l_{2_2},\dots,l_{2_t},\dots,l_{z_1},l_{z_2},\dots,l_{z_t})\;\neq\\(a_{2_1},a_{2_2},\dots,a_{2_t},\dots,a_{z_1},a_{z_2},\dots,a_{z_t})}}\{C_{2,l_{2_1}}\cup C_{2,l_{2_2}}\cup\dots\cup C_{z,l_{z_t}}\}\right\}\right\}$$

	$$f_m = \;C_{1,e_1}\bigcap\left\{\bigcap_{\substack{l_{s_k}\;=\;\{i_{s_0},i_{s_1},\dots,i_{s_t}\},\\l_{s_k}\;\neq\;l_{s_{k'}},\\\forall k,\;k'\in[1,t]\text{ and }\forall s \;\in\; [2,z] \\ (l_{2_1},l_{2_2},\dots,l_{2_t},\dots,l_{z_1},l_{z_2},\dots,l_{z_t})\;\neq\\(a_{2_1},a_{2_2},\dots,a_{2_t},\dots,a_{z_1},a_{z_2},\dots,a_{z_t})}}\{C_{2,l_{2_1}}\cup C_{2,l_{2_2}}\cup\dots\cup C_{z,l_{z_t}}\}\right\}$$
	
	Proceeding in a similar manner, we can write
	
	$$f_m = \;\{C_{1,e_1}\cap C_{2,e_2}\}\bigcap\\
	\left\{\bigcap_{\substack{l_{s_k}\;=\;\{i_{s_0},i_{s_1},\dots,i_{s_t}\},\\l_{s_k}\;\neq\;l_{s_{k'}},\\\forall k,\;k'\in[1,t]\text{ and }\forall s \;\in\; [2,z] \\ (l_{3_1},l_{3_2},\dots,l_{3_t},\dots,l_{z_1},l_{z_2},\dots,l_{z_t})\;\neq\\(a_{3_1},a_{3_2},\dots,a_{3_t},\dots,a_{z_1},a_{z_2},\dots,a_{z_t})}}\{C_{3,l_{3_1}}\cup C_{3,l_{3_2}}\cup\dots\cup C_{z,l_{z_t}}\}\right\}$$
	
	$$f_m = \;\{C_{1,e_1}\cap C_{2,e_2}\cap C_{3,e_3}\}\bigcap\\
	\left\{\bigcap_{\substack{l_{s_k}\;=\;\{i_{s_0},i_{s_1},\dots,i_{s_t}\},\\l_{s_k}\;\neq\;l_{s_{k'}},\\\forall k,\;k'\in[1,t]\text{ and }\forall s \;\in\; [2,z] \\ (l_{4_1},l_{4_2},\dots,l_{4_t},\dots,l_{z_1},l_{z_2},\dots,l_{z_t})\;\neq\\(a_{4_1},a_{4_2},\dots,a_{4_t},\dots,a_{z_1},a_{z_2},\dots,a_{z_t})}}\{C_{4,l_{4_1}}\cup C_{4,l_{4_2}}\cup\dots\cup C_{z,l_{z_t}}\}\right\}$$
	
	$$\vdots$$

	$$f_m = \;\{C_{1,e_1}\cap C_{2,e_2}\cap\dots\cap C_{z-1,e_{z-1}}\}\bigcap\\
	\left\{\bigcap_{\substack{l_{z_k}\;=\;\{i_{z_0}, i_{z_1},\dots,i_{z_t}\},\\(l_{z_1},l_{z_2},\dots,l_{z_t})\;\neq\;\\(a_{z_1},a_{z_2},\dots,a_{z_t})}}\{C_{z,l_{z_1}}\cup C_{2,l_{z_2}}\cup\dots\cup C_{z,l_{z_t}}\}\right\}$$
	
	$$f_m = \;\{C_{1,e_1}\cap C_{2,e_2}\cap\dots\cap C_{z,e_{z}}\}$$
\end{IEEEproof}

\begin{lem} At the end of each transmission corresponding to set $\mathcal{X}$, each user is able to decode one sub-file.
\end{lem}
\begin{IEEEproof}
	Consider a user $m,\;m\in \mathcal{X}$ and any user $m',\;m'\neq m,\;m'\in\mathcal{X}$. The set $f_{m'}$ represents the set of subfiles which is available with every  user in $\mathcal{X}$, other than $m'$. The user $m$ has access to all the subfiles in $f_{m'}$. Consider the transmission corresponding to set $\mathcal{X}$
	$$\underset {m \in \mathcal{X}} \oplus W_{d_{m},y_{m,s}},\; s \in [\mu_{z}]$$
	The $m^{th}$ user is able to get the subfile $ W_{d_{m},y_{m,s}}$
	from this transmission since  it has every other subfile $W_{d_{m'},y_{m',s}},\;\forall m'\in \mathcal{X},\;m'\neq m$.
\end{IEEEproof}	

\begin{lem} At the end of all the transmissions in {\bf Algorithm 1} each user is able get all subfiles of the demanded file $W_{d_m},\;m\in \mathcal{K}.$ 
\end{lem}

\begin{IEEEproof}
	In {\bf Algorithm 1} it can noted that there are in total $\binom{(b_r-1)^z}{t}$ transmissions from which the $m^{th}$ user gets $\mu_z\binom{(b_r-1)^z}{t}$ subfiles. Now consider a combination of $(t+1)z$ blocks (caches) given as $C_{1,i_{1_0}},C_{1,i_{1_1}},\dots,C_{1,i_{1_t}}$, $C_{2,i_{2_0}},C_{2,i_{2_1}},\dots,C_{2,i_{2_t}}$, $\dots,C_{z,i_{z_0}},C_{z,i_{z_1}},\dots,C_{z,i_{z_t}}$,
	where $i_{s_k} \in [1,b_r]$  and  $i_{s_k} \neq i_{s_{k'}},\;\forall k,\;k' \in [0,t]$ and $\forall s \in [1,z]$, and the $m^{th}$ user having access to $tz$ blocks (caches), $C_{1,a_{1_1}},C_{1,a_{1_2}},\dots,C_{1,a_{1_t}}$, $C_{2,a_{2_1}},C_{2,a_{2_2}},\dots,C_{2,a_{2_t}}$, $\dots,C_{z,a_{z_1}}C_{z,a_{z_2}},\dots, C_{z,a_{z_t}}$, where $a_{s_k}\in \{i_{s_0},i_{s_1},\dots,i_{s_t}\}$ and  $a_{s_k}\neq a_{s_{k'}},\;\forall k,\;k' \in [1,t]$ and $\forall s = [1,z]$. We have
	
	$$Y_m = C_{1,a_{1_1}}\cup C_{1,a_{1_2}}\cup\dots\cup C_{1,a_{1_t}}\cup\dots\cup C_{z,a_{1_1}}\cup C_{z,a_{z_2}}\cup\dots\cup C_{z,a_{z_t}}$$
	
	and the $m^{th}$ user is able to receive subfile indices of the demanded file $W_{d_m}$ from the transmission corresponding to $(t+1)z$ caches (considered above) given by
	
	$$f_m = \bigcap_{\substack{
			l_{s_k}\;=\;\{i_{s_0},i_{s_1},\dots,i_{s_t}\},\\l_{s_k}\;\neq\;l_{s_{k'}},\\\forall k,\;k'\in[1,t]\text{ and }\forall s \;\in\; [1,z] \\ (l_{1_1},l_{1_2},\dots,l_{1_t},\dots,l_{z_1},l_{z_2},\dots,l_{z_t})\;\neq\\(a_{1_1},a_{1_2},\dots,a_{1_t},\dots,a_{z_1},a_{z_2},\dots,a_{z_t})}}\{C_{1,l_{1_1}}\cup C_{1,l_{1_2}}\cup\dots\cup C_{z,l_{1_t}}\cup\dots\cup C_{z,l_{z_1}}\cup C_{z,l_{z_2}}\cup\dots\cup C_{z,l_{z_t}}\}$$
	
	which, using Lemma 2 is same as $\{C_{1,e_1}\cap C_{2,e_2}\cap\dots\cap C_{z,e_{z}}\}$ where, $e_s = \{i_{s_0},i_{s_1},\dots,i_{s_t}\}\setminus\{a_{s_1},a_{s_2},\dots,a_{s_t}\},\;\forall s\in [1,z]$. In order to find subfile indices that $m^{th}$ user get from all $\binom{(b_r-1)^z}{t}$ transmissions, we have to vary the value of $e_s$ such that $e_s\neq a_s,\;\forall s\in [z]$. The $m^{th}$ user is able to receive subfile indices given by
	
	$$\bigcup_{\substack{e_s\;=\;1,\\e_s\;\notin\;\{a_{s_1},a_{s_2},\dots,a_{s_t}\}\\\forall s\;\in\; [z]}}^{b_r}\{C_{1,e_1}\cap C_{2,e_2}\cap\dots\cap C_{z,e_{z}}\}.$$
	
	In addition to this using $Y_m,$ the $m^{th}$ user gets the subfile indices shown in the sequence of expressions shown below
	
	$$
	=Y_m\bigcup\left\{\bigcup_{\substack{e_s\;=\;1,\\e_s\;\notin\;\{a_{s_1},a_{s_2},\dots,a_{s_t}\}\\\forall s\;\in\; [z]}}^{b_r}\{C_{1,e_1}\cap C_{2,e_2}\cap\dots\cap C_{z,e_{z}}\}\right\}
	$$
	
	\begin{multline*}
	=\left\{C_{1,a_{1_1}}\cup C_{1,a_{1_2}}\cup\dots\cup C_{1,a_{1_t}}\cup\dots\cup C_{z,a_{z_1}}\cup C_{z,a_{z_2}}\cup\dots\cup C_{z,a_{z_t}}\right\}\bigcup\\
	\bigcup_{\substack{e_s\;=\;1,\\e_s\;\notin\;\{a_{s_1},a_{s_2},\dots,a_{s_t}\}\\\forall s\;\in\; [z-1]}}^{b_r}\left\{C_{1,e_1}\cap C_{2,e_2}\cap\dots\cap \left\{\overset{b_r}{\underset{\substack{e_z=1\\e_z\;\notin\;\{a_{z_1},a_{z_2},\dots,a_{z_t}\}}}{\cup}}C_{z,e_{z}}\right\}\right\}
	\end{multline*}
	
	\begin{multline*}
	\supseteq\left\{C_{1,a_{1_1}}\cup C_{1,a_{1_2}}\cup\dots\cup C_{1,a_{1_t}}\cup\dots\cup C_{z-1,a_{{z-1}_1}}\cup C_{z-1,a_{{z-1}_2}}\cup\dots\cup C_{z-1,a_{{z-1}_t}}\right\}\bigcup\\\bigcup_{\substack{e_s\;=\;1,\\e_s\;\notin\;\{a_{s_1},a_{s_2},\dots,a_{s_t}\}\\\forall s\;\in\; [z-1]}}^{b_r}\left\{C_{1,e_1}\cap C_{2,e_2}\cap\dots\cap  \{C_{z,a_{z_1}}\cup C_{z,a_{z_2}}\cup\dots\cup C_{z,a_{z_t}}\}\right\}\bigcup\\	\bigcup_{\substack{e_s\;=\;1,\\e_s\;\notin\;\{a_{s_1},a_{s_2},\dots,a_{s_t}\}\\\forall s\;\in\; [z-1]}}^{b_r}\left\{C_{1,e_1}\cap C_{2,e_2}\cap\dots\cap \left\{\overset{b_r}{\underset{\substack{e_z=1\\e_z\;\notin\;\{a_{z_1},a_{z_2},\dots,a_{z_t}\}}}{\cup}}C_{z,e_{z}}\right\}\right\}
	\end{multline*}
	
	\begin{multline*}
	=\left\{C_{1,a_{1_1}}\cup C_{1,a_{1_2}}\cup\dots\cup C_{1,a_{1_t}}\cup\dots\cup C_{z-1,a_{{z-1}_1}}\cup C_{z-1,a_{{z-1}_2}}\cup\dots\cup C_{z-1,a_{{z-1}_t}}\right\}\bigcup\\
	\bigcup_{\substack{e_s\;=\;1,\\e_s\;\notin\;\{a_{s_1},a_{s_2},\dots,a_{s_t}\}\\\forall s\;\in\; [z-1]}}^{b_r}\left\{C_{1,e_1}\cap C_{2,e_2}\cap\dots\cap\left\{\overset{b_r}{\underset{e_z=1}{\cup}}C_{z,e_{z}}\right\}\right\}
	\end{multline*}
	
	\begin{multline*}
	=\left\{C_{1,a_{1_1}}\cup C_{1,a_{1_2}}\cup\dots\cup C_{1,a_{1_t}}\cup\dots\cup C_{z-1,a_{{z-1}_1}}\cup C_{z-1,a_{{z-1}_2}}\cup\dots\cup C_{z-1,a_{{z-1}_t}}\right\}\bigcup\\
	\bigcup_{\substack{e_s\;=\;1,\\e_s\;\notin\;\{a_{s_1},a_{s_2},\dots,a_{s_t}\}\\\forall s\;\in\; [z-1]}}^{b_r}\left\{C_{1,e_1}\cap C_{2,e_2}\cap\dots\cap C_{z-1,e_{z-1}}\right\}
	\end{multline*}
	
	Following the similar steps, we have\\
	
	\begin{multline*}
	\supseteq\left\{C_{1,a_{1_1}}\cup C_{1,a_{1_2}}\cup\dots\cup C_{1,a_{1_t}}\cup\dots\cup C_{z-2,a_{{z-2}_1}}\cup C_{z-2,a_{{z-2}_2}}\cup\dots\cup C_{z-2,a_{{z-2}_t}}\right\}\bigcup\\
	\bigcup_{\substack{e_s\;=\;1,\\e_s\;\notin\;\{a_{s_1},a_{s_2},\dots,a_{s_t}\}\\\forall s\;\in\; [z-2]}}^{b_r}\left\{C_{1,e_1}\cap C_{2,e_2}\cap\dots\cap C_{z-2,e_{z-2}}\right\}
	\end{multline*}
	
	$$\vdots$$
	
	$$\supseteq\left\{C_{1,a_{1_1}}\cup C_{1,a_{1_2}}\cup\dots\cup C_{1,a_{1_t}}\cup C_{2,a_{2_1}}\cup C_{2,a_{2_2}}\cup\dots\cup C_{2,a_{2_t}}\right\}\bigcup
	\left\{\bigcup_{\substack{e_s\;=\;1,\\e_s\;\notin\;\{a_{s_1},a_{s_2},\dots,a_{s_t}\}\\\forall s\;\in\; [2]}}^{b_r}\left\{C_{1,e_1}\cap C_{2,e_2}\right\}\right\}$$

	$$\supseteq\left\{C_{1,a_{1_1}}\cup C_{1,a_{1_2}}\cup\dots\cup C_{1,a_{1_t}}\right\}\bigcup
	\left\{\bigcup_{\substack{e_1\;=\;1,\\e_1\;\notin\;\{a_{1_1},a_{1_2},\dots,a_{1_t}\}}}^{b_r}C_{1,e_1}\right\}$$
	
	$$ = \bigcup_{\substack{e_1\;=\;1}}^{b_r}C_{1,e_1}$$
	
	The above expression is the union of all the blocks in a parallel class. So from the property of resolvable designs  the above set is equal to the set containing all the subfile indices of all the files and therefore it also contains all the subfiles of file $W_{d_m}$.
\end{IEEEproof}

\end{document}